\documentclass{article}
\usepackage{spconf,amsmath,graphicx,amssymb,hyperref,url,multirow,caption,subcaption,changepage}
\usepackage{arydshln}
\usepackage[pagewise,modulo]{lineno}
\usepackage[labelsep=period]{caption}

\title{Music Mixing Style Transfer: A Contrastive Learning Approach \\to Disentangle Audio Effects}

\name{\parbox{\linewidth}{\centering$^*$Junghyun Koo$^{1, 3}$ \qquad Marco A. Martínez-Ramírez$^{1}$ \qquad Wei-Hsiang Liao$^{1}$
\\ \qquad Stefan Uhlich$^{2}$ \qquad Kyogu Lee$^{3}$ \qquad Yuki Mitsufuji$^{1}$}
\thanks{*This work was done during an internship at Sony. \newline contact: dg22302@snu.ac.kr}
}

\address{$^{1}$ Sony Group Corporation, Tokyo, Japan \qquad$^{2}$ Sony Europe B.V., Stuttgart, Germany \\ $^{3}$ Music and Audio Research Group, Department of Intelligence and Information, Seoul National University }

\begin{document}
\ninept
\maketitle
\begin{abstract}
We propose an end-to-end music mixing style transfer system that converts the mixing style of an input multitrack to that of a reference song. This is achieved with an encoder pre-trained with a contrastive objective to extract only audio effects related information from a reference music recording. 
All our models are trained in a self-supervised manner from an already-processed wet multitrack dataset with an effective data preprocessing method that alleviates the data scarcity of obtaining unprocessed dry data.
We analyze the proposed encoder for the disentanglement capability of audio effects and also validate its performance for mixing style transfer through both objective and subjective evaluations.
From the results, we show the proposed system not only converts the mixing style of multitrack audio close to a reference but is also robust with mixture-wise style transfer upon using a music source separation model.
\end{abstract}


\vspace{-5pt}
\section{Introduction}
\label{sec:intro}
\vspace{-5pt}
Due to the subjective nature of personal music taste, ideal intelligent music production systems should produce results reckoning the users' preferences \cite{moffat2019approaches}, since musicians and audio engineers often get their inspiration from reference music. 
However, replicating a musical \textit{audio effects} (FX) style is usually time-consuming and difficult to do, especially in post-production tasks such as music mixing, which requires significant training to provide users with a natural and pleasant listening experience. 
As such, there is a need to develop music systems that can imitate the style of reference music, to make the music production process more efficient \cite{key}.
Recent studies of intelligent music production have utilized neural networks for style transfer of FX with differentiable \cite{lee2022differentiable, ramirez2021differentiable, steinmetz2022style} or black-box approaches \cite{koo2021reverb, koo2022end}.
Yet, these approaches focus on a few specific FX and lack evaluation of style transfer on multitrack recordings.
To this end, we aim to transfer the mixing style of input music stems to sound as the desired track.

We first propose a novel approach to encoding FX using an encoder trained in a self-supervised manner with a contrastive objective. Then we perform music mixing style transfer from the encoded representation of the reference track to make input stems have a similar mixing style to that of the reference. 
Our evaluations mainly focus on the performance of the encoder and show it successfully extracts FX from music recordings and further validate its impact on mixing style transfer.

Detailed implementation of our system are publicly available\footnote{\url{https://github.com/jhtonyKoo/music_mixing_style_transfer}}, and the supplementary results and generated audio samples are presented on our demo page: \url{https://jhtonyKoo.github.io/MixingStyleTransfer/}

\vspace{-5pt}
\section{Methodology}
\label{sec:methodology}
\begin{figure*}[t]
  \centering
  \includegraphics[width=\linewidth]{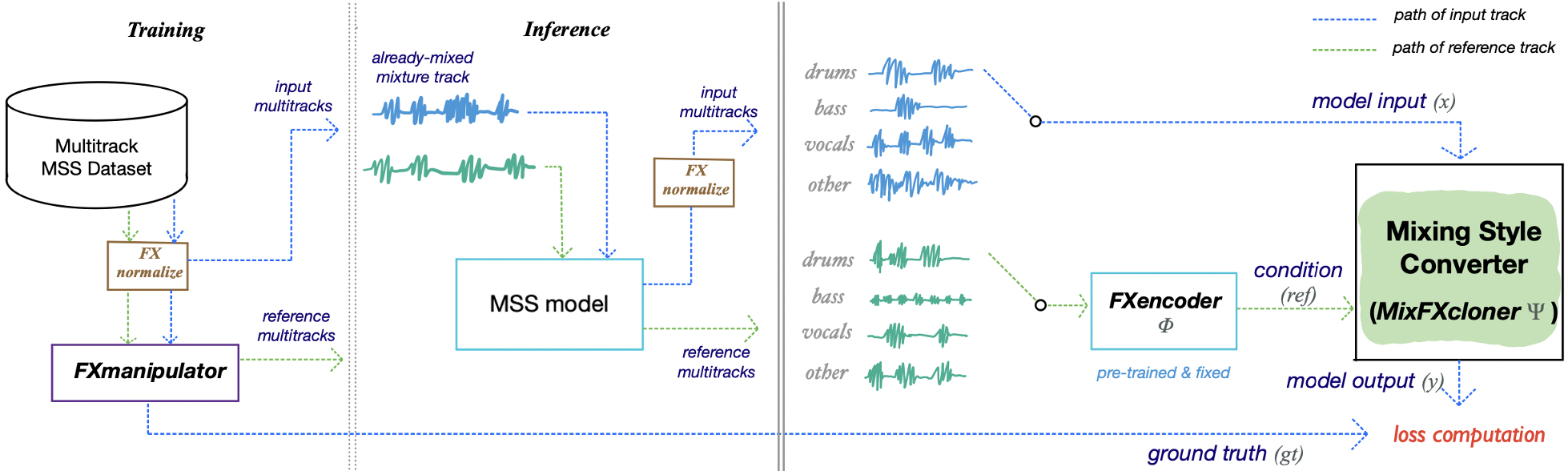}
  \setlength{\abovecaptionskip}{-10pt}
  \caption{Overview of the proposed music mixing style transfer system.
  \textit{MixFXcloner} $\Psi$ aims to convert the mixing style of the input track to that of the reference track. This is assisted by \textit{FXencoder} $\Phi$ which was pre-trained with a contrastive objective to encode only the audio effects-related information of the given audio track. 
  During the training stage, we randomly select two different segments of each instrument track from the Music Source Separation (MSS) dataset then \textit{FX normalize} them. We input one of the chosen segments to the \textit{MixFXcloner}, where its objective is to match the mixing style of the other track manipulated by the \textit{FXmanipulator}. We apply the same manipulation to the input to produce a ground truth sample for loss computation.
  During the inference stage, not only we can transfer the mixing style of single-stem inputs, but also mixture-wise inputs by using an MSS model.
}
  \label{fig:model_pipeline}
\vspace{-15pt}
\end{figure*}

\vspace{-5pt}
The proposed system is trained in a self-supervised manner, where we use a public available Music Source Separation (MSS) dataset to train each model. Fig.\ref{fig:model_pipeline} depicts the overall pipeline of the proposed system, which we will now explain in detail.

\subsection{Audio Effects Chain Manipulation}
\label{ssec:fx_chain_manipulation}

Our audio effects manipulator (\textit{FXmanipulator}) is a chain of 6 different fundamental FX commonly used during music mixing, and its purpose is to generate random mixing styles for the self-supervised training scheme of our system.
The order of applying effects is 1. \textit{equalization} (EQ), 2. \textit{dynamic range compression} (DRC), 3. \textit{panning}, 4. \textit{stereo imaging}, 5. \textit{convolution reverberation}, and 6. \textit{loudness gain}, where we also randomly shuffle the order of EQ and DRC, and panning and imaging during manipulation.
Each FX is applied with random parameters according to a predefined probability which allows the \textit{FXmanipulator} to produce a larger variation of FX.

\vspace{-5pt}
\subsubsection{FX normalization}
\vspace{-3pt}
A mixing style transfer system requires a method of generating the same style of FX from a reference with different musical content. Such a system could be trained in a supervised manner by manipulating both input and reference mixes given that dry or clean recordings are available.
However, most publicly available music data are already processed with certain FX, limiting the accessibility of training data.
An alternative way to generate a target mix with the FX style of the reference is to apply the \textit{FX normalization} data preprocessing method proposed in \cite{martinez2022automatic} to the wet recordings.
This diminishes the variation of FX and thus allows any wet data to be similarly transformed into a homogeneous style.
For example, if we normalized the loudness of two different songs to $-10dB$ LUFS, we can expect the normalized songs to have almost the same volume value before applying our \textit{FXmanipulator}. As shown in \cite{martinez2022automatic}, we follow the same principle for the different types of applied FX.
Note that the normalized or homogeneous mixing style would inevitably introduce variance to the resulting FX characteristics due to the inherent variation of the effects-normalization methods.
However, we remark on the importance of \textit{FX normalization} in Sec. \ref{sec:experiments}.

The procedure of \textit{FX normalization} is done in the following order: EQ, DRC, stereo imaging, panning, and loudness gain. We follow the same normalization process\footnote{\url{https://github.com/sony/fxnorm-automix}} of \cite{martinez2022automatic} except for imaging and panning, where we perform these normalizations not according to frequency bins but in the time domain.
Thus, \textit{FX normalization} effectively restricts the default FX distribution, allowing more variation while being accurate when normalized stems are randomly manipulated using the \textit{FXmanipulator}.

\subsection{Audio Effects Encoder}
\label{ssec:fx_encoder}
The main idea of the Audio Effects Encoder (\textit{FXencoder}) $\Phi$ is to disentangle FX information from music recordings, which is difficult to accomplish due to the inherent interrelation between musical timbre and audio effects \cite{bromham2019impact}.
We adopt a contrastive learning approach to solely discriminate the difference of FX among various mixing styles.
The original idea of contrastive learning aims to encode a robust characteristic of the content \cite{chen2020simple}, where \cite{spijkervet2021contrastive,koo2022end} applied this idea to the music domain for encoding the overall identity of songs.
We redesign its intention to make the encoder focus on distinguishing only the manipulation of audio effects.

To train \textit{FXencoder} on accurate FX guidance, we first pre-process the input data with \textit{FX normalization} to normalize features related to audio effects.
From these normalized samples, we apply the same manipulation using \textit{FXmanipulator} to two different contents and assign them as positive pairs. A batch of manipulated tracks is formed by repeating this procedure, where contents applied with different manipulations are assigned as negative samples. Every content is manipulated twice with different configurations of the \textit{FXmanipulator}, which helps the model to disentangle FX and content by ensuring a sample with the same content but a different style among negative samples.
During training, different combinations of mixes are created while ensuring that both positive and negative mixes contain the same type of stems.

We further introduce a training strategy called \textit{probability scheduling}, which adjusts the \textit{FXencoder} to balance out the importance of each FX.
If the model is trained only with samples applied with a full set of FX, the model may only be attentive to easily distinguishable FX (e.g., panning) for segregation, which leads to a less informative representation on minor FX.
Therefore, instead of applying a fixed set of probabilities, we reschedule them in the range of [0.1, 1.0] according to each validation loss value computed with samples only applied with every single FX to apply easier or more difficult FX less or more often, respectively.
This rescheduling process is possible since the objective function of the \textit{FXencoder}, normalized temperature-scaled cross-entropy loss introduced in SimCLR, evaluates regardless of the acoustic impact caused by each FX manipulation.

The architecture of the \textit{FXencoder} is the same as the Music Effects Encoder (MEE) proposed in \cite{koo2022end} which is composed of multiple residual 1-dimensional convolutional blocks. The model is capable of encoding variable length of stereo audio input producing a single 2048-dimensional representation vector.
Following the training procedure of \textit{DirectCLR} \cite{jing2021understanding}, 
we remove the projection layers and directly backpropagate subvectors of the representation vector $\Phi(.)[0:d_{0}]$, where $d_0$ is a hyperparameter. This technique makes sure that the model fully uses the encoded representation for the mixing style transfer task and prevents dimensional collapse, i.e., not being able to utilize the entire embedding space.

\begin{table*}[t]
\centering
\caption{Model Description and Objective Measure of FX Encoders}
\label{table:obj_measure_encoder}
\setlength\tabcolsep{3.5pt}
\footnotesize
\begin{tabular}{c:cc|cc:cc:cc:cc}
\hline
\multirow{2}{*}{\textbf{Model}} & \multirow{2}{*}{\textit{\begin{tabular}[c]{@{}c@{}}audio effects\\ normalization\end{tabular}}} & \multirow{2}{*}{\textit{\begin{tabular}[c]{@{}c@{}}probability\\ scheduling\end{tabular}}} & \multicolumn{2}{c:}{\textbf{multitrack. w/ full FX}} & \multicolumn{2}{c:}{\textbf{single stem. w/ full FX}} & \multicolumn{2}{c:}{\textbf{multitrack. w/ single FX}} & \multicolumn{2}{c}{\textbf{single stem. w/ single FX}} \\
                                &                                                                                      &                                                                                            & \textbf{DCIMIG}       & \textbf{DCI RF Expl}         & \textbf{DCIMIG}           & \textbf{DCI RF Expl}      & \textbf{DCIMIG}            & \textbf{DCI RF Expl}      & \textbf{DCIMIG}            & \textbf{DCI RF Expl}      \\ \hline
MEE \cite{koo2022end}                             & -                                                                                    & -                                                                                          & 0.129                 & 0.816 ± 0.004                & 0.131             & 0.802 ± 0.027             & 0.074              & 0.727 ± 0.079             & 0.064              & 0.655 ± 0.104             \\
$\Phi_{\varnothing}$                         & -                                                                                    & -                                                                                          & 0.398                 & 0.928 ± 0.000                & 0.371             & 0.910 ± 0.031             & 0.209              & 0.830 ± 0.074             & 0.124              & 0.772 ± 0.081             \\
$\Phi_{\text{norm}}$                         & $\checkmark$                                                                         & -                                                                                          & 0.473                 & 0.937 ± 0.001                & 0.430             & \textbf{0.932 ± 0.028}    & 0.248              & 0.836 ± 0.092             & 0.146              & 0.786 ± 0.093             \\
$\Phi_{\text{p.s.}}$                         & $\checkmark$                                                                         & $\checkmark$                                                                               & \textbf{0.674}        & \textbf{0.963 ± 0.000}       & \textbf{0.515}    & 0.928 ± 0.023             & \textbf{0.349}     & \textbf{0.881 ± 0.059}    & \textbf{0.183}     & \textbf{0.795 ± 0.059}    \\ \hline
\end{tabular}
\vspace{-15pt}
\end{table*}

\subsection{Music Mixing Style Converter}
\label{ssec:style_converter}
The proposed mixing style converter (\textit{MixFXcloner}) $\Psi$ transfers the mixing style of the reference track \textit{ref} to the input audio $x$ to produce the output mixture $y$. 
\textit{MixFXcloner} converts the FX style of a single stem at a time.
Therefore, the model performs a multitrack conversion by processing each stem individually.
The training is carried out by computing the difference between $y$ and the ground truth $gt$, which is generated by applying to $x$ the same \textit{FXmanipulator} that has been applied to \textit{ref}.

In this work, we mainly focus on the impact of different configurations of the \textit{FXencoder} and, hence, we use a single configuration of \textit{MixFXcloner} to evaluate the performance of music mixing style transfer. We adopt a temporal convolutional network (TCN) \cite{steinmetz2022efficient} with a receptive field of 5.2 seconds. The encoded embedding of the reference track $\Phi(\textit{ref})$ is conditioned with a feature-wise linear modulation (FiLM) \cite{perez2018film} operation at each TCN block.

The objective function of the \textit{MixFXcloner} is multi-scale spectral loss $\mathcal{L}_{\text{MSS}}$ \cite{engel2020ddsp} on both left-right and mid-side channels (mid = left + right, side = left - right).
We utilize the original loss function with FFT sizes of (4096, 2048, 1024, 512) where each spectral loss is
$\mathcal{L}_i = (1-\alpha)||S_i-\hat{S}_i||_1 + \alpha||\log S_i-\log\hat{S}_i||_2^2$. $S_i$ and $\hat{S}_i$ denotes magnitude spectrogram of $gt$ and $y$ with the $i$\textsuperscript{th} FFT size, respectively, and $\alpha$ is set to $0.1$ in all our experiments. The final loss function for \textit{MixFXcloner}, where $\mathcal{L}_{\text{MSS}} = \sum_i \mathcal{L}_i $, is then:
\begin{equation}
\begin{split}
    \mathcal{L}_{\Psi} &= \mathcal{L}_{\text{MSS}}(gt_{\text{left}}, y_{\text{left}}) + \mathcal{L}_{\text{MSS}}(gt_{\text{right}}, y_{\text{right}}) \\ 
    &+ \mathcal{L}_{\text{MSS}}(gt_{\text{mid}}, y_{\text{mid}})
    + \mathcal{L}_{\text{MSS}}(gt_{\text{side}}, y_{\text{side}}) .
\end{split}
\end{equation}
\vspace{-10pt}

\begin{figure}[t]
     \centering
     \setlength{\abovecaptionskip}{5.0pt}
     \begin{subfigure}[b]{0.1175\textwidth}
         \centering
         \includegraphics[width=\textwidth]{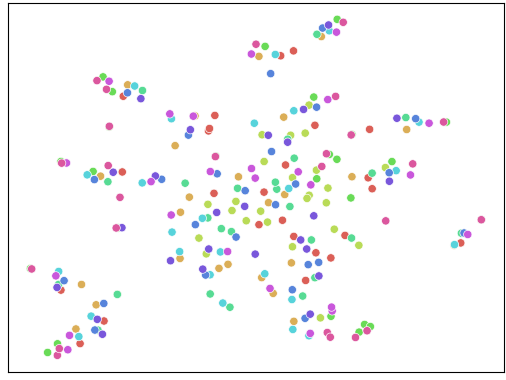}
         \setlength{\abovecaptionskip}{-8pt}
         \caption{MEE}
         \label{fig:t-sne_fullfx-MEE}
     \end{subfigure}
     \begin{subfigure}[b]{0.1175\textwidth}
         \centering
         \includegraphics[width=\textwidth]{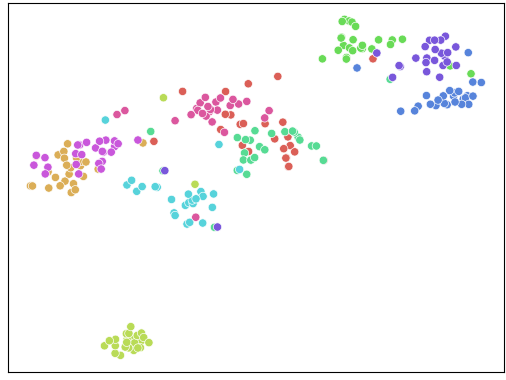}
         \setlength{\abovecaptionskip}{-8pt}
         \caption{$\Phi_{\varnothing}$}
         \label{fig:t-sne_fullfx-model1}
     \end{subfigure}
     \begin{subfigure}[b]{0.1175\textwidth}
         \centering
         \includegraphics[width=\textwidth]{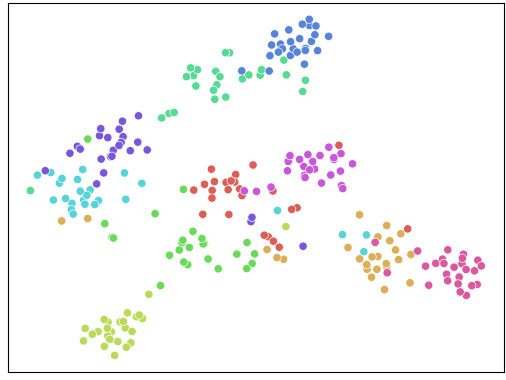}
         \setlength{\abovecaptionskip}{-8pt}
         \caption{$\Phi_{\text{norm}}$}
         \label{fig:t-sne_fullfx-model2}
     \end{subfigure}
     \begin{subfigure}[b]{0.1175\textwidth}
         \centering
         \includegraphics[width=\textwidth]{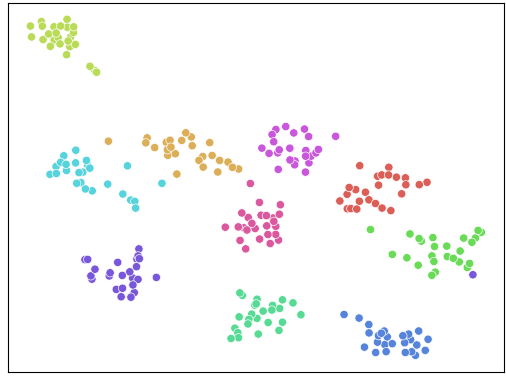}
         \setlength{\abovecaptionskip}{-8pt}
         \caption{$\Phi_{\text{p.s.}}$}
         \label{fig:t-sne_fullfx-model3}
     \end{subfigure}
    \\
     \begin{subfigure}[b]{0.1175\textwidth}
         \centering
         \includegraphics[width=\textwidth]{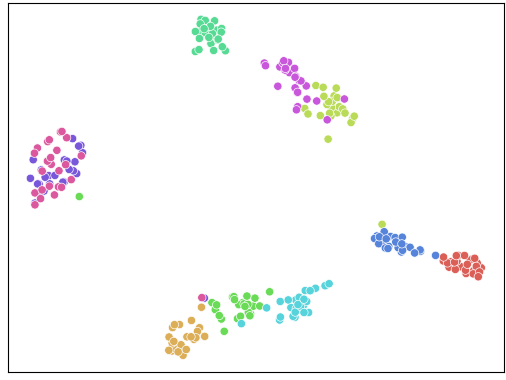}
         \setlength{\abovecaptionskip}{-8pt}
         \caption{FX: panning - $\Phi_{\text{norm}}$}
         \label{fig:t-sne_pan-model2}
     \end{subfigure}
     \begin{subfigure}[b]{0.1175\textwidth}
         \centering
         \includegraphics[width=\textwidth]{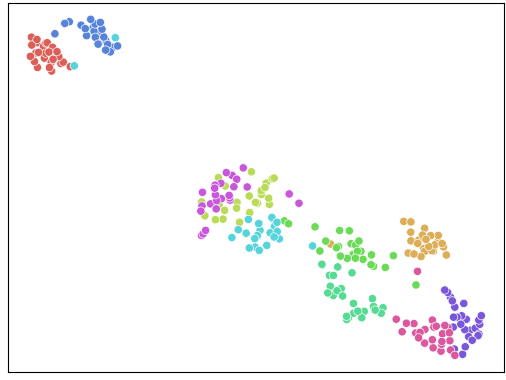}
         \setlength{\abovecaptionskip}{-8pt}
         \caption{FX: panning - $\Phi_{\text{p.s.}}$}
         \label{fig:t-sne_pan-model3}
     \end{subfigure}
     \begin{subfigure}[b]{0.1175\textwidth}
         \centering
         \includegraphics[width=\textwidth]{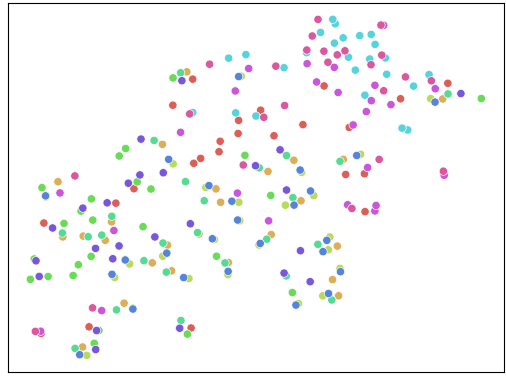}
         \setlength{\abovecaptionskip}{-8pt}
         \caption{FX: stereo imaging - $\Phi_{\text{norm}}$}
         \label{fig:t-sne_imaging-model2}
     \end{subfigure}
     \begin{subfigure}[b]{0.1175\textwidth}
         \centering
         \includegraphics[width=\textwidth]{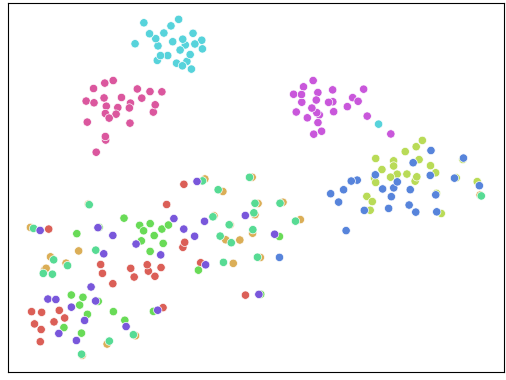}
         \setlength{\abovecaptionskip}{-8pt}
         \caption{FX: stereo imaging - $\Phi_{\text{p.s.}}$}
         \label{fig:t-sne_imaging-model3}
     \end{subfigure}
        \caption{t-SNE samples of embeddings from FX Encoders - multitrack applied with full \textit{(a)$\sim$(d)} or a single \textit{(e)$\sim$(h)} FX manipulation}
        \label{fig:t-SNE_multitrack_fullsinglefx}
    \vspace{-15pt}
\end{figure}

\section{Experiments}
\label{sec:experiments}
\vspace{-5pt}
This work mainly assesses \textit{FXencoder} with its effectiveness in extracting FX from a music recording and how informative the encoded representation is to performing mixing style transfer.
In this section, we compare our method with MEE, a model pre-trained with a self-supervised representation learning approach to extract music mastering style information, which shares a similar approach of encoding FX of a high-fidelity music piece.
To observe the impact of audio effects normalization and probability scheduling, we compare 3 different \textit{FXencoders} listed in Table \ref{table:obj_measure_encoder}.
We first perform qualitative and quantitative evaluation on the disentangled representation of FX, then evaluate the performance of \textit{MixFXcloner} trained with these encoders for mixing style transfer objectively and subjectively.

\subsection{Dataset}
\label{ssec:dataset}
All our models are trained and validated using the MUSDB18 dataset \cite{rafii2017musdb18}, a publicly available dataset widely used for the MSS task. The dataset consists of a total of 150 songs, where each song is a multitrack of four different instruments: drums, bass, vocals, and other. We split the dataset into 86, 14, and 50 songs for training, validation, and test, respectively.
Based on \cite{martinez2022automatic}, we precompute the effects characteristics of each instrument from the training subset, then apply \textit{FX normalization} using these features to the entire dataset. 
Silent regions of the training and validation sets are removed since the FX manipulation applied on silence will likely cause inaccurate objectives for the system.

\subsection{Experimental Setup}
\label{ssec:exp_setup}
Following the training procedure of MEE \cite{koo2022end}, we incorporate input audio with a variable duration of $[5.9, 11.8]$ seconds. 
All of our \textit{FXencoders} are trained with a batch size of $256$ for $16$K iterations, where an iteration consists of 4 steps of encoding $[1 .. 4]$ different combinations of multitrack mix. We use the Adam optimizer \cite{kingma2014adam} with a learning rate of $2\cdot10^{-4}$ and it is scheduled with a cosine annealing \cite{loshchilov2016sgdr} without restart. The temperature parameter for the InfoNCE loss is set to $0.1$ and we use $d_0$ $360$ which was the best performing subset dimensionality in \cite{jing2021understanding}. For the evaluation of MEE, we use the pre-trained model from the original repository\footnote{\url{https://github.com/jhtonyKoo/e2e_music_remastering_system}}.

We train \textit{MixFXcloner} with a batch size of 8 for 100 epochs, where an epoch is defined as $min($total duration of each instrument track$) \div len(x)$. The duration of $x$ and \textit{ref} are both $11.8$ seconds for both training and validation. The same optimizer and learning rate used to train \textit{FXencoders} are applied.

\subsection{Disentangled Representation of the \textit{FXencoder}}
\label{ssec:eval_enc}
We evaluate the representation of the \textit{FXencoder} by latent-space visualizations and with objective metrics that measure disentanglement of FX. 
For the evaluation, we augment 25 normalized segments of multitracks randomly selected from the test subset and we manipulate each stem individually with 10 different random configurations of \textit{FXmanipulator}.

\subsubsection{Qualitative Evaluation}
\label{ssec:t-sne}
Fig. \ref{fig:t-SNE_multitrack_fullsinglefx} shows the t-SNE scatter plots of each embedding from different encoders.
Points with the same color represent segments of different songs manipulated with the same \textit{FXmanipulator} parameters. Therefore, each color group contains the same 25 songs.
From Fig.~\ref{fig:t-sne_fullfx-MEE}, we observe that MEE is not capable of disentangling FX from the content although it was trained to identify the overall timbre of songs.
On the other hand, all of our three models formed clusters according to each group of FX. 
Yet, we see that the clusters from the $\Phi_{\varnothing}$ are more entangled to each other, which is the outcome of the model being trained with initial samples of large FX variance, causing an inaccurate objective after FX manipulation.

Fig. \ref{fig:t-sne_pan-model2}---\ref{fig:t-sne_imaging-model3} demonstrates the importance of our proposed probability scheduling. Instead of applying the full FX chain, we apply a single FX at a time to explore the encoded representation in detail. 
Both $\Phi_{\text{norm}}$ and $\Phi_{\text{p.s.}}$ form reasonable clusters upon panning, while $\Phi_{\text{norm}}$ lacks its disentanglement upon the stereo imager FX.
This can be explained due to the encoder being trained with a fixed set of FX probabilities, thus it focuses more on the dominant FX while failing to encode less dominant FX.
On the contrary, $\Phi_{\text{p.s.}}$ is better at clustering on stereo imaging since FX probability scheduling allowed the encoder to successfully train on manipulated samples with respect to less dominant FX.
The full t-SNE results can be found on our demo page.

\subsubsection{Objective Evaluation}
\label{eval_obj_enc}
To quantitatively measure the representation of \textit{FXencoder}, we adopt DCIMIG \cite{sepliarskaia2019evaluating} and \textit{Disentangle Metric for Informativeness} (DCI RF Expl) \cite{eastwood2018framework} as disentanglement metrics.
DCIMIG is a metric that evaluates overall disentanglement properties, and DCI RF Expl measures the explicitness: a critical factor for our downstream task of style transfer that represents completeness of the representation of describing factors of interest.
We follow the implementation of \cite{carbonneau2022measuring}, where all output values of each metric are normalized in the range of $[0.0, 1.0]$.
Table \ref{table:obj_measure_encoder} summarizes disentanglement performance on representation obtained from combinations of multiple/single FX applied on multitrack/single stem.
Results are averaged over each test case, where we compute DCI RF Expl with four different random seeds for each case to also obtain the standard deviation.
As observed from Fig. \ref{fig:t-SNE_multitrack_fullsinglefx}, all of our proposed encoders outperform MEE. Additionally, we observe the increasing performance according to each proposed training technique except for encoding the case of full FX applied on a single instrument, where $\Phi_{\text{norm}}$ is slightly better on DCI RF Expl than $\Phi_{\text{p.s.}}$.
This can be inferred as an aftermath of probability scheduling where $\Phi_{\text{norm}}$ could be trained to discriminate samples by focusing only on the dominant FX, whereas $\Phi_{\text{p.s.}}$ further shows its strength on single FX cases.

{\renewcommand{\arraystretch}{1.2}
\begin{table}[t]
\caption{Objective Measure of Mixing Style Transfer: values indicate conversion results of multitrack / single stem evaluation}
\label{table:obj_measure_style_transfer}
\begin{adjustwidth}{-0.1cm}{}
\setlength\tabcolsep{3.2pt}
\fontsize{8}{8}\selectfont
\begin{tabular}{c|cccc}
\hline
\multirow{2}{*}{\textbf{Method}} & \multicolumn{2}{c}{$\mathbf{\mathcal{L}_{\text{MSS}}}$} & \multicolumn{2}{c}{$\mathbf{\lambda\cdot\Delta \Phi_{\text{p.s.}}}$}                                                                          \\
                                 & \textbf{left\&right}         & \textbf{mid\&side}           & \textbf{$\mathbf{y}$ - $\mathbf{ref}$} & \textbf{$\mathbf{y}$ - $\mathbf{gt}$} \\ \hline
\multicolumn{1}{l|}{Input}                            & 1.222 / 0.679                & 1.620 / 0.897                & 0.405 / 0.222                                                          & 0.359 / 0.203                                                        \\ \hdashline
\multicolumn{1}{l|}{$\Psi$ w/ MEE}                           & 1.198 / 0.675                & 1.533 / 0.844                & 0.400 / 0.213                                                          & 0.346 / 0.187                                                        \\
\multicolumn{1}{l|}{$\Psi$ w/ $\Phi_{\varnothing}$}                       & 1.114 / 0.610                & 1.431 / 0.770                & 0.306 / 0.168                                                          & 0.254 / 0.141                                                        \\
\multicolumn{1}{l|}{$\Psi$ w/ $\Phi_{\text{norm}}$}                       & \textbf{0.980 / 0.542}       & 1.309 / 0.718                & 0.294 / 0.163                                                          & 0.237 / 0.135                                                        \\
\multicolumn{1}{l|}{$\Psi$ w/ $\Phi_{\text{p.s.}}$}                       & 0.982 / 0.543                & \textbf{1.274 / 0.688}       & \textbf{0.251 / 0.136}                                                 & \textbf{0.180 / 0.107}                                               \\ \hline
seperated stems                    & 2.250 / 1.470                & 2.994 / 1.875                & 0.245 / 0.112                                                          & 0.225 / 0.172                                                        \\ \hline
\end{tabular}
\end{adjustwidth}
\vspace{-15pt}
\end{table}
}

\vspace{-3pt}
\subsection{Mixing Style Transfer}
\label{ssec:eval_style_transfer}

\subsubsection{Objective Evaluation}
\label{ssec:eval_obj_style}

To measure the performance of the FX style transfer, we use $\lambda\cdot\Delta \Phi_{\text{p.s.}}$ which computes the $\mathcal{L}1$ distance between encoded embeddings from $\Phi_{\text{p.s.}}$, where the term $\lambda=10^3$ for scaling.
The average $\lambda\cdot\Delta \Phi_{\text{p.s.}}$ between $gt$ and \textit{ref} is 0.214 and 0.119 for mixture and stem-wise results, respectively.
(Note that $\Phi_{\text{p.s.}}$ of mixture is not $\frac{1}{k} \sum_{i=1}^k \Phi_{\text{p.s.}}($stem$_i)$ but $\Phi_{\text{p.s.}}(\sum_{i=1}^k$stem$_i)$.)
According to the results of Table \ref{table:obj_measure_style_transfer}, \textit{MixFXcloner} trained with $\Phi_{\text{p.s.}}$ as \textit{FXencoder} is the best performing method except for $\mathcal{L}_{\text{MSS}}$ on left and right channels with a minor difference compared to $\Phi_{\text{norm}}$. Overall, we observe the tendency of each model's performance to be the same as in Sec. \ref{ssec:eval_enc}, which implies that the representation from the encoder fully reflects the performance of style transfer.

To evaluate the performance upon the inference pipeline, we source separated both the summation of input stems $\sum x$ and target stems $\sum \textit{ref}$ with Hybrid Demucs \cite{defossez2021hybrid}, 
an open-source MSS model trained with the training subset of MUSDB18
that achieved high performance on a realistic dataset in the Music Demixing Challenge \cite{mitsufuji2021music},
then used them as inputs to \textit{MixFXcloner} trained with $\Phi_{\text{p.s.}}$.
The mean SDR value between the original and source separated outputs are 8.004 and 4.554 for mixture and stem-wise conversion, respectively. The higher SDR for the mixture could be due to the fact that the artifacts caused from the MSS model get cancelled out on mixture level conversion. 
Moreover, using separated stems show almost the same performance on $\Delta \Phi_{\text{p.s.}}$ as clean ones, indicating that the conversion capacity of the system is robust to artifacts.

\subsubsection{Subjective Evaluation}
\label{ssec:eval_sjb_style}
Although the objective measurements of style transfer may convey the actual perceptual performance thanks to the presence of a ground truth, perceptual assessment is further required to validate the objective findings.
Hence, a test was conducted with the Web Audio Evaluation Tool \cite{waet2015} and had been aimed at professional audio engineers only. Participants were asked to rate different mixes based on their similarity to the reference mix in terms of FX characteristics and mixing style.
Each test question consists of two reference tracks ($x$ and \textit{ref}) and five different stimuli tracks.
We designed the test with a total of 12 questions consisting of 4 questions of full multitrack conversion and 2 questions for each instrument of single stem conversion.

\begin{figure}[t]
     \centering
     \setlength{\abovecaptionskip}{5.0pt}
     \begin{subfigure}[b]{0.23\textwidth}
         \centering
         \includegraphics[width=\textwidth]{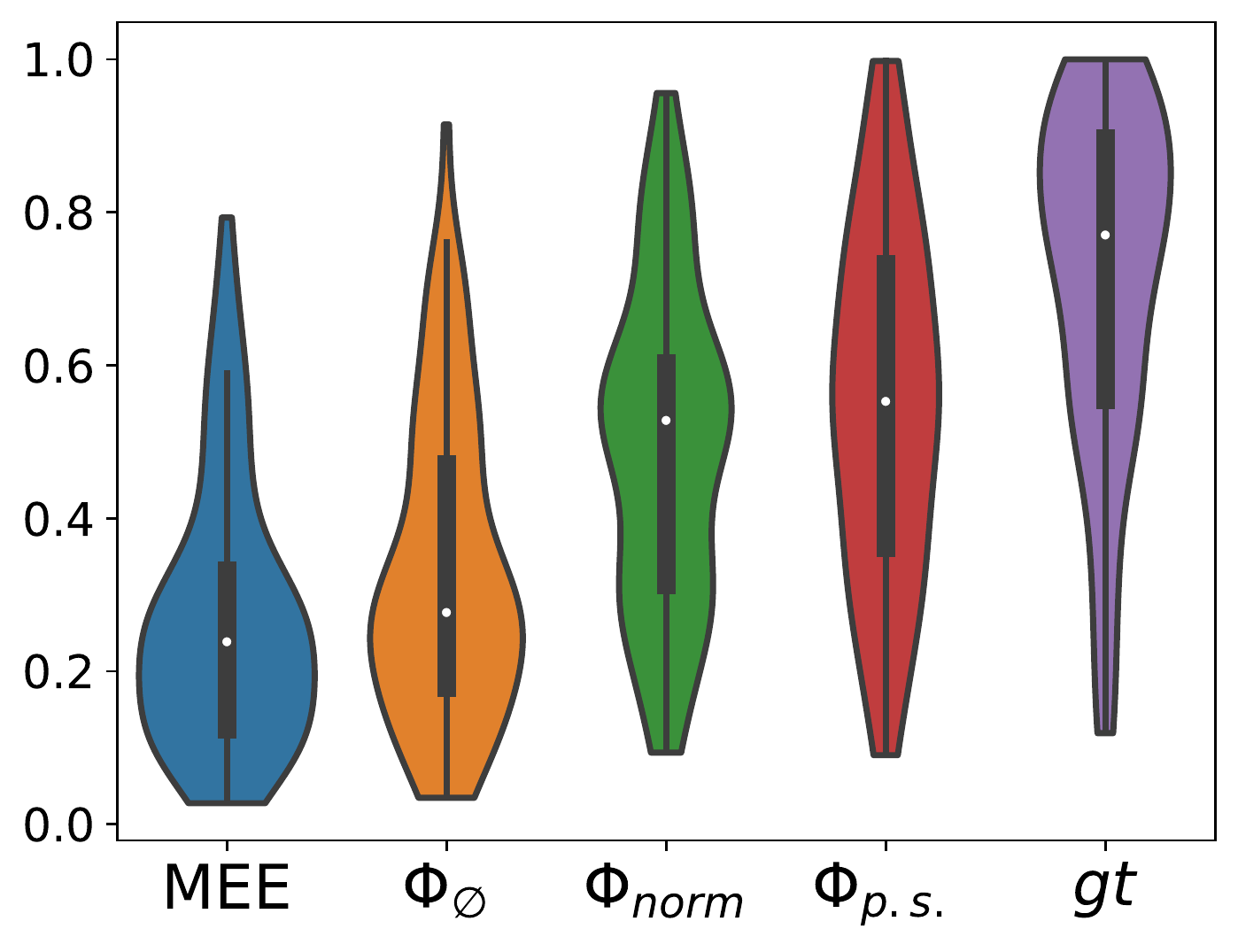}
         \setlength{\abovecaptionskip}{-8pt}
         \caption{single stem conversion}
         \label{fig:single_stem_conversion}
     \end{subfigure}
     \begin{subfigure}[b]{0.23\textwidth}
         \centering
         \includegraphics[width=\textwidth]{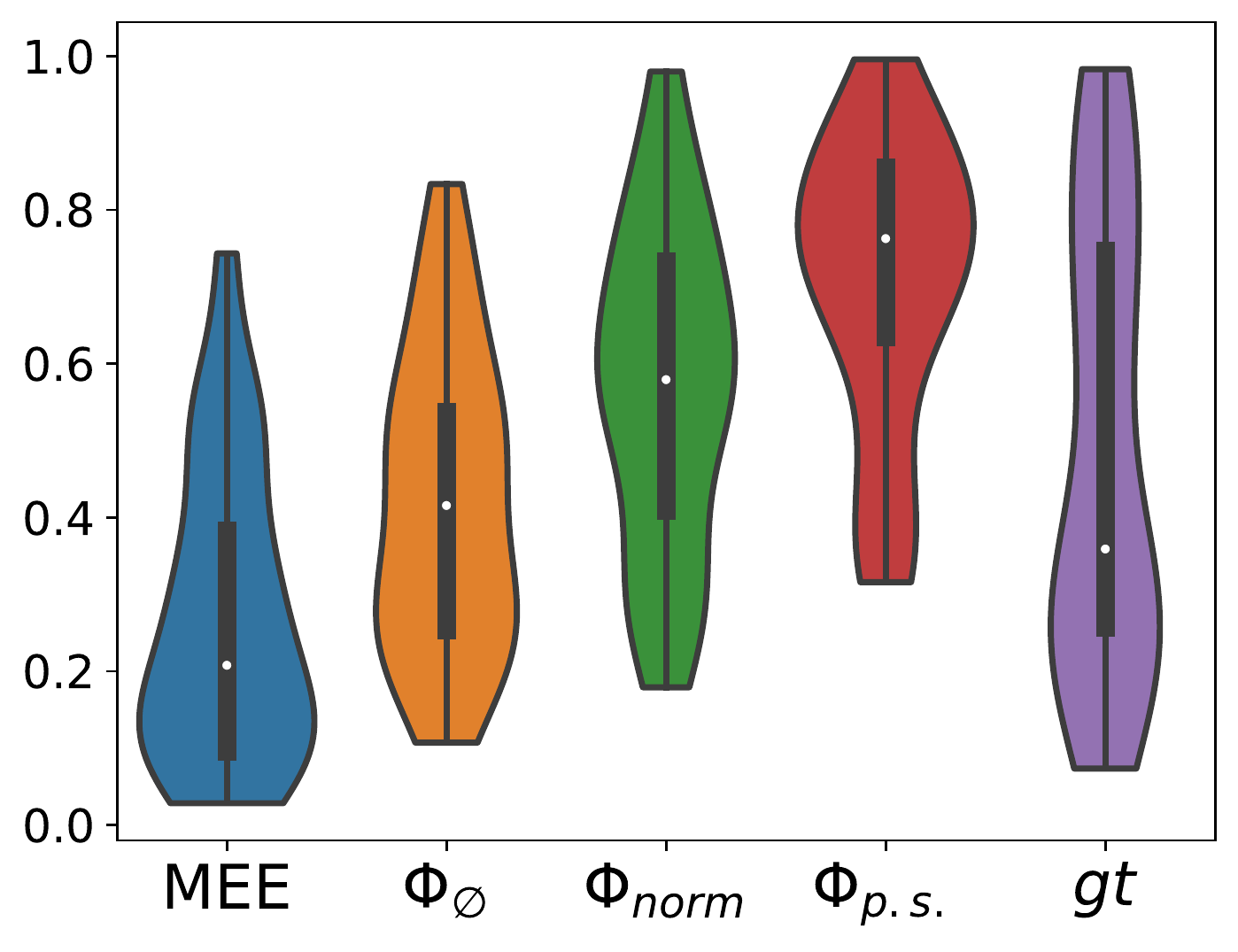}
         \setlength{\abovecaptionskip}{-8pt}
         \caption{multitrack conversion}
         \label{fig:multitrack_conversion}
     \end{subfigure}
    \caption{Listening test violin plots}
    \label{fig:listening_test}
    \vspace{-15pt}
\end{figure}

The results of single stem and full multitrack mixing style conversion are summarized in Fig. \ref{fig:listening_test}. In total, 11 audio engineers with an average mixing experience of 7.3 years participated in the test.
Based on the analysis of post-hoc paired t-tests with Bonferroni correction, there was no significant difference between $gt$ and $\Phi_{\text{norm}}$ and $\Phi_{\text{p.s.}}$ for multitrack conversion, and for single stem conversion, MEE and $\Phi_{\varnothing}$, and $\Phi_{\text{norm}}$ and $\Phi_{\text{p.s.}}$. All other results showed a significant difference with $p<0.05$.
The result of single stem conversion highlights the importance of \textit{FX normalization} where \textit{MixFXcloner} trained with $\Phi_{\varnothing}$ is not rated hight than the model trained with MEE. Although there was no significant difference between $\Phi_{\text{norm}}$ and $\Phi_{\text{p.s.}}$ for the single stem conversion, we observe that $\Phi_{\text{p.s.}}$ outperforms all other methods on multitrack conversion.

Overall, similar performance was observed compared to the objective metrics, with the exception of the $gt$ ratings in the multitrack style conversion. We attribute this phenomenon to the increased difficulty inherent in rating a full multitrack mixing style conversion task.
Upon comparing styles of a single stem, participants can precisely focus on each applied FX and make the comparison more accurate.
On the contrary, evaluating multitrack conversion makes the comparison much more complicated, since the complexity of the different combinations of FX together with the different contents between $x$ and \textit{ref} drastically increases the difficulty of the task even for professional audio engineers.
Also, given the intrinsic variance of the \textit{FX normalization} procedure, it is worth noting that $gt$ and \textit{ref} will not always be a perfect match in terms of FX characteristics. Therefore, we hypothesize that this is a phenomenon that can be seen in Fig. \ref{fig:listening_test}, although further analysis is needed.

\section{Conclusion}
\label{sec:conclusion}
\vspace{-5pt}
We propose a novel contrastive learning approach to encode FX information from single or multitrack music data. We then train an end-to-end mixing style transfer system, where the FX characteristics of a reference track are successfully encoded and used to condition a converter to perform FX style transfer to a given set of input stems.
Our system is trained in a self-supervised manner where no dry music recordings are used during training. 
We evaluate our system both with objective and subjective measures and show that \textit{FXencoder} disentangles audio effects from a music recording and its capacity is highly correlated to the performance of \textit{MixFXcloner}.

\vfill\pagebreak

\bibliographystyle{IEEEbib}
\bibliography{refs}

\end{document}